\documentclass[%
preprint,
superscriptaddress,
 amsmath,amssymb,
 aps,
]{revtex4-1}

\usepackage{bm}
\usepackage{ulem}
\usepackage{cancel}

\usepackage{physics}

\newcommand{\e}{\mathrm{e}}
\begin{document}



\title{
Screened and Unscreened Solutions for Relativistic Star \\
 in de Rham-Gabadadze-Tolley (dRGT) Massive Gravity
}

\author{Masashi~Yamazaki}
\email{yamazaki.masashi@c.mbox.nagoya-u.ac.jp}
\affiliation{Department of Physics, Nagoya University, Nagoya 464-8602, Japan}
\author{Taishi~Katsuragawa}
\email{taishi@mail.ccnu.edu.cn}
\affiliation{Institute of Astrophysics, Central China Normal University, Wuhan 430079, China}
\author{Sergei~D.~Odintsov}
\email{odintsov@ieec.uab.es}
\affiliation{ICREA, Passeig Luis Companys, 23, 08010 Barcelona, Spain}
\affiliation{Institute of Space Sciences (IEEC-CSIC) C.\@ Can Magrans s/n, 08193 Barcelona, Spain}
\author{Shin'ichi~Nojiri}
\email{nojiri@gravity.phys.nagoya-u.ac.jp}
\affiliation{Department of Physics, Nagoya University, Nagoya 464-8602, Japan}
\affiliation{Kobayashi-Maskawa Institute for the Origin of Particles and the Universe, 
Nagoya University, Nagoya 464-8602, Japan}


\begin{abstract}
We study the static and spherical symmetric (SSS) configurations
in the non-minimal model of the de Rham-Gabadadze-Tolley (dRGT) massive gravity 
with a flat reference metric.
Considering the modified Tolman-Oppenheimer-Volkoff (TOV) equation, the Bianchi identity, 
and energy-momentum conservation,
we find a new algebraic equation for the radial coordinate of the reference metric.
We demonstrate that this equation suggests an absence of the Vainshtein mechanism 
in the minimal model of the dRGT massive gravity,
while it has two branches of solutions where one connects with the Schwarzschild space-time 
and another implies the significant deviation from the asymptotically flat space-time 
in the non-minimal model.
We also briefly discuss the boundary conditions for the relativistic stars in the dRGT massive gravity
and a potential relation with the mass-radius relation of the stars.
\end{abstract}

\maketitle


\section{Introduction}
\label{sec_intro}

One of the aims of modified gravity theories is to explain phenomena 
which is hardly understood in the framework of the general relativity.
Several modified gravity theories are motivated to study the dark energy related to 
the accelerated expansion of the 
Universe~\cite{Riess1998,Perlmutter1999,Spergel2003}.
The cosmological constant $\Lambda$ gives us a simple solution to the dark energy problem, 
where $\Lambda$ may be interpreted as the vacuum energy induced from 
the quantum fluctuation of matter fields. 
However, it suffers from two theoretical problems: the fine-tuning problem and 
coincidence problem (for example, see \cite{Weinberg1989a, Padilla2015a}).

To explain the late-time acceleration of our Universe without invoking the cosmological constant, 
one needs to introduce the long-distance (IR) modifications of gravity theories
so that the modification is responsible for the cosmic acceleration at present.
On the other hand, such modifications often bring us the unsuitable feature which is to be 
excluded by the observations.
It is well known that the Solar-System observations are consistent with the prediction 
in the general relativity (see \cite{Debono2016} for a review),  and thus, 
the IR modification should be hidden in such a situation.

The modification for the dark energy is often regarded as the dynamical dark energy 
characterized by additional fields.
Thus, if such dynamical fields induced from the IR modification are suppressed on local scales, 
one can safely avoid the constraint from the observations in the Solar-System.
The screening mechanisms~\cite{Vainshtein1972a, Khoury2004, doi:10.1142/S0218271809016107} 
suggest the way for making the additional degrees of freedom ineffective in short-distance.
Therefore, the viable modified gravity theories should possess the screening mechanism, and then, 
they do not conflict with Solar-System constraints, keeping the validity to introduce the IR 
modification for the dark energy problem.

Although plenty of the previous research has verified the screening mechanisms in the static 
and spherical symmetric (SSS) configurations,
the screening mechanisms are not well-understood in the highly dense matter region to study 
the effect of the modification on the short-distance behavior.
A typical situation can be found in the relativistic star.
The hydrostatic equilibrium, which is maintained in the balance between the pressure of internal 
matters and its gravity, determines the inner and outer structures of relativistic stars.
It means that the series of mass and radius of relativistic stars depends on 
the models of hadron physics in the high-density matter and gravitational theories 
in the strong-gravity region.
 From the perspective of hadron physics, various equations of state (EoS) have been 
investigated~\cite{doi:10.1146/annurev-nucl-102711-095018}, corresponding to the 
inner structure of relativistic stars as in \cite{Weber2005}. 
The EoS determines the mass-radius relation of relativistic stars and the maximum mass, 
and the existence of massive neutron star with the mass larger than $2M_\odot$ is, 
at present, one of the criteria for the realistic model of EoS.~\cite{Demorest2010}.

 From another point of view, the gravity theories also determine the mass-radius relations,
where the behavior in the non-perturbative or non-linear region is of great importance 
to the inner structure of the relativistic stars.
In the previous works~\cite{Astashenok2013, Astashenok2014, Capozziello2016},
one of the authors has applied $F(R)$ gravity theories to 
the study on the relativistic stars.
Because the curvature of space-time, the Ricci scalar $R$, around relativistic stars is larger 
than that in the Solar-System,
some models of $F(R)$ gravity show significant differences from general relativity ($F(R)=R$) 
around the relativistic stars.
These notable results imply that 
the relativistic stars can be a useful tool for investigating the modifications of gravity.
Furthermore, the study on the relativistic stars also provides us with a better understanding 
of screening mechanisms.
Several works have attempted to study the Vainshtein mechanism~\cite{Vainshtein1972a}, 
which is one of the screening mechanism.
By assuming a constant-density profile inside the star, the Vainshtein mechanism has 
been discussed ~\cite{Volkov2013} in the presence of matter fields.

In this work, we study the relativistic stars in the de Rham-Gabadadze-Tolley (dRGT) 
massive gravity~\cite{DeRham2011b},
which is the theory of a ghost-free massive spin-2 particle. 
The theory has five ghost-free modes
that are two tensor modes, two vector modes, and a scalar mode.
The vector modes cannot couple with matters because of the energy-momentum conservation,
while the additional scalar mode can produce different matter coupling with gravity from that 
in the general relativity.
This additional scalar mode is considered to be suppressed by the Vainshtein mechanism;
the non-linear derivative couplings hide the scalar mode, and the gravitational coupling with 
matters become similar to that in the general relativity inside so-called the Vainshtein radius.

In our previous work~\cite{Katsuragawa2016a},
we studied the relativistic stars in the minimal model of dRGT massive gravity for the SSS 
configuration with flat reference metric and found that the maximum mass of relativistic 
stars become smaller than that of general relativity.
In the light of this results,
we postulate that the lack of the Vainshtein mechanism results in the smaller maximal mass, due to 
the absence of non-linear kinetic couplings in the relativistic stars scales.
A theoretical analysis for the minimal model shows that the minimal model does not have 
the Vainshtein mechanism~\cite{Renaux-Petel2014a}.

The purpose of this article is 
to study the system of the relativistic stars in the non-minimal model, which is 
the broader framework of dRGT massive gravity, and to determine how 
the Vainshtein mechanism would affect the mass-radius relation of the relativistic star.
We will derive the modified Tolman-Oppenheimer-Volkoff (TOV) equations
to see how the modifications of gravity change the inside and outside structures of 
the relativistic star
After that, we will discuss the behavior of the solutions of modified TOV equations
according to their mathematical structure,
to find that the system of interest has a solution which is very similar to that in 
the general relativity thanks to the non-linear kinetic terms.
This study provides new insights into the non-perturbative aspects dRGT massive gravity, 
and we argue that the Vainshtein mechanism potentially works around the relativistic star.

\section{Modified TOV Equation in dRGT Massive Gravity}
\subsection{dRGT Massive Gravity}
\label{sec_2A}

In this section, 
we derive equations of motion of dRGT massive gravity in the SSS configuration and 
show the modified TOV equations.
In the units of $c=\hbar=1$, the action of the dRGT massive gravity~\cite{Hassan2011} 
can be written as
\begin{equation}
S_\mathrm{dRGT}
=\frac{1}{2\kappa^2}\int d ^4x
\sqrt{-\det(g)}
\left[ R(g) -2m_0^2\sum_{n=0}^3\beta_n e_n\left(\sqrt{g^{-1}f}\right) \right] + S_\mathrm{mat}
\, ,
\label{action}
\end{equation}
where $\kappa^2=8\pi G$ is the gravitational coupling constant and $S_\mathrm{mat}$ is 
the matter action.
We are using the units of $c=\hbar=1$.
The $e_k(\mathbb{X})$ are polynomials defined as the anti-symmetric products of the components
\begin{equation}
e_k\pqty{\mathbf{X}}
 =\frac{1}{k!}\mathbf{X}^{I_1}{}_{[I_1}\cdots\mathbf{X}^{I_k}{}_{I_k]}
\, .
\label{interaction1}
\end{equation}
The action \eqref{action} includes the two metric tensors $g_{\mu \nu}$ and $f_{\mu \nu}$; 
$g_{\mu \nu}$ denotes the dynamical variable in the dRGT massive gravity
while $f_{\mu \nu}$ is fixed by hand and called as the reference or fiducial metric.
The $\sqrt{g^{-1}f}$ represents the matrix such that
\begin{equation}
 \pqty{\sqrt{g^{-1}f}}^\mu{}_\rho
 \pqty{\sqrt{g^{-1}f}}^\rho{}_\nu
 =g^{\mu\rho}f_{\rho\nu}
\, .
\label{square-root_of_metric}
\end{equation}
Here, $m_0$ is a parameter which defines the graviton mass, and in the following analysis, 
we set it as
\begin{align}
m_0\equiv 10^{-33}\mathrm{eV}\sim \pqty{10^{26}\mathrm{m}}^{-1}
\, .
\label{graviton_mass1}
\end{align}
This value is the same order of the cosmological constant, 
which represents the IR modification for the dark energy, 
and it is consistent with several observations (see for a review~\cite{DeRham2016a}).
The parameters $\beta_n$'s are free and expressed by only two parameters
if we demand the flat Minkowski space-time as a solution of the field equations
and the appropriate coefficient of the graviton-mass term as in the Fierz-Pauli 
theory~\cite{Hassan2012}:
\begin{equation}
\beta_0=6-4\bar{\alpha}_3+\bar{\alpha}_4\, , \quad 
\beta_1=-3+3\bar{\alpha}_3-\bar{\alpha}_4\, , \quad 
\beta_2=1-2\bar{\alpha}_3+\bar{\alpha}_4\, , \quad 
\beta_3=\bar{\alpha}_3-\bar{\alpha}_4\, ,
\label{beta1}
\end{equation}
and they lead to the algebraic relations between parameters $\beta_{n}$:
\begin{equation}
\beta_2=1-\beta_0-2\beta_1 \, ,\quad 
\beta_3=-3+2\beta_0+3\beta_1\, .
\label{beta2}
\end{equation}
We call the case that
\begin{equation}
\beta_0=3\, ,\quad 
\beta_1=-1\, ,\quad 
\beta_2=\beta_3=0\, ,
\label{minimal_model}
\end{equation}
as the minimal model~\cite{Hassan2012}
by meaning the minimal non-linear extension of Fierz-Pauli theory~\cite{Fierz1939a}.
In this work, we restrict our discussion for the case that $\beta_n\sim\order{1}$;
otherwise, conditions for the UV completion are violated~\cite{Cheung2016}.

By the variation of action \eqref{action} with respect to the dynamical metric $g_{\mu \nu}$,
we obtain the equations of motion in the dRGT massive gravity as follows (for the derivation, 
see \cite{Katsuragawa2016a}):
\begin{equation}
G_{\mu\nu} + m_0^2 I_{\mu\nu} = \kappa^2 T_{\mu\nu}
\, ,
\label{eom}
\end{equation}
where
\begin{align}
I_{\mu \nu}\equiv& \sum^{3}_{n=0}(-1)^{n}\beta_{n} 
g_{\mu \lambda}Y^{\lambda}_{(n) \nu}(\sqrt{g^{-1}f}) 
\, ,
\label{interaction2}
\\
Y^{\lambda}_{(n) \nu}(\mathbf{X})\equiv & \sum^{n}_{r=0} (-1)^{r} \left( X^{n-r}
\right)^{\lambda}_{\ \nu} e_{r}(\mathbf{X})
\, .
\label{Yint1}
\end{align}
Here, the matrix $Y_{(n)}(\mathbf{X})$ are written in the following forms:
\begin{align}
Y_{0}(\mathbf{X}) =& \mathbf{1}\, , \quad
Y_{1}(\mathbf{X}) = \mathbf{X}- \mathbf{1}[\mathbf{X}]\, , \quad
Y_{2}(\mathbf{X}) = \mathbf{X}^{2} - \mathbf{X}[\mathbf{X}]
+ \frac{1}{2} \mathbf{1} \left( [\mathbf{X}]^{2} - [\mathbf{X}^{2}]
\right)
\, , \nonumber \\
Y_{3}(\mathbf{X}) =& \mathbf{X}^{3} - \mathbf{X}^{2}[\mathbf{X}]
+ \frac{1}{2} \mathbf{X} \left( [\mathbf{X}]^{2} - [\mathbf{X}^{2}] \right)
- \frac{1}{6} \mathbf{1} \left( [\mathbf{X}]^{3} -
3[\mathbf{X}][\mathbf{X}^{2}] + 2[\mathbf{X}^{3}] \right) 
\, .
\label{Yint2}
\end{align}
Now, we have to pay attention to the lack of diffeomorphism invariance because of the existence 
of graviton mass.
While the diffeomorphism invariance guarantees the universal graviton-matter coupling 
in the general relativity, 
we should assume the universal couplings, which ensures to eliminate the ghost modes 
in the dRGT massive gravity.

\subsection{Ansatz for SSS Configuration}
\label{sec_2B}

Considering a relativistic star,
we impose the SSS configuration to $g_{\mu\nu}$ metric.
Then, the ansatz for $g_{\mu \nu}$ can be written as,
\begin{align}
\label{physical_metric}
g_{\mu\nu} d  x^\mu  d  x^\nu
= -\e^{2\nu(r)} d  t^2 + \e^{2\lambda(r)}  d  r^2 + r^2 d \Omega^2 
\, .
\end{align}
$\nu(r)$ and $\lambda(r)$ are functions with respect to $r$,
and $\e^{2\lambda(r)}$ is related to the mass function in the ordinary TOV equation:
\begin{align}
\label{mass_function}
\e^{-2\lambda(r)} \equiv 1-\frac{2GM(r)}{r}
\, ,
\end{align}
where $M(r)$ is the mass parameter.
The functions $\nu(r)$ and $\lambda(r)$ should satisfy boundary conditions that
they vanish at the center of the star
\begin{equation}
\label{boundary_condition1}
\nu(r=0) = \lambda(r=0) = 0
\, ,
\end{equation}
which suggests that the conical singularity should be removed~\cite{wald1984general}.
The boundary conditions indicate 
that the mass parameter $M(r)$ should also vanish at the center,
\begin{equation}
\label{boundary_condition2}
M(r=0) = 0
\, .
\end{equation}
The equation of motion Eq.~\eqref{eom} determine the asymptotic behavior of these 
three functions.
Note that the space-time around the SSS configurations asymptotically matches 
with the Minkowski space-time in the general relativity.

We assume the $f_{\mu\nu}$ metric as follows in our model:
\begin{equation}
\label{reference_metric}
f_{\mu\nu} d  x^\mu  d  x^\nu = - d  t^2 +  d \chi(r)^2 + \chi(r)^2 d \Omega^2 
= - d  t^2 + \chi^{\prime}(r)^2 d  r^2 + \chi(r)^2 d \Omega^2 \, ,
\end{equation}
where the prime denote the derivative with respect to $r$.
The reference metric $f_{\mu \nu}$ is chosen to represent the flat space-time,
while the radial coordinate is, in general, different from that of the physical metric $g_{\mu \nu}$.
The relationship of the radial coordinate between $g_{\mu\nu}$ and $f_{\mu\nu}$ space-time 
is reflected to the new function $\chi(r)$.
That is, if $\chi(r)=r$, 
$f_{\mu \nu}$ describes exactly the Minkowski space-time from the observer 
in the coordinate system of $g_{\mu \nu}$.
The function $\chi(r)$ is determined by new algebraic equations derived from 
the divergence of equations of motion.
In the later section, we will show the new algebraic equation is at most fourth order.

\subsection{Modified TOV equation}
\label{sec_2C}

Substituting the ansatz with respect to the dynamical and reference metrics, 
$g_{\mu \nu}$ and $f_{\mu \nu}$, 
into the equations of motion \eqref{eom}, 
we obtain the modified TOV equations:
\begin{align}
\label{eq:eom_t}
-\kappa^2\rho 
&= -\frac{1}{r^2} + \frac{1}{r^2}\left(1 +r\partial_r\right)\e^{-2\lambda}
+ m_0^2 I^t{}_t 
\, ,\\
\label{eq:eom_r}
\kappa^2 p 
&= -\frac{1}{r^2}+\frac{1+2r\nu'}{r^2}\e^{-2\lambda}+m_0^2I^r{}_r 
\, ,\\
\label{eom_theta}
\kappa^2 p
&= \pqty{\nu''+\nu'^2+\frac{\nu'-\lambda'}{r}-\nu'\lambda'}\e^{-2\lambda}
+m_0^2I^\theta_\theta
\, .
\end{align}
And the conservation law of the energy-momentum tensor leads to
\begin{align}
\label{conservation}
-\frac{p'}{p+\rho}&=\nu'
\, .
\end{align}
Rewriting the above four equations Eqs.~\eqref{eq:eom_t}, \eqref{eq:eom_r}, \eqref{eom_theta}, 
and \eqref{conservation},
we obtain the following expressions 
\begin{align}
GM'&=4\pi G\rho r^2+\frac{1}{2}m_0^2r^2I^t{}_t
\label{eq:mtov_mass}
\, , \\
-\frac{p'}{p+\rho}&=\nu'
=\frac{4\pi Gpr^3+GM-\frac{1}{2}m_0^2r^3I^r{}_r}{r(r-2GM)}
\label{eq:mtov_pres}
\, , \\
\kappa^2 p
&=\pqty{\nu''+\nu'^2+\frac{\nu'}{r}}\pqty{1-\frac{2GM}{r}}
+\frac{1}{2}\pqty{\frac{1}{r}+\nu'}\pqty{1-\frac{2GM}{r}}'
+m_0^2I^\theta{}_\theta \, .
\label{eq:mtov_theta}
\end{align}
One can find that the original TOV equations in the general relativity are modified by 
the interaction terms $m^{2}_{0} I^{i}_{\ i}$, where $i=t, r, \theta$.
We can compute the modification terms $I^{t}_{\ t}$, $I^{r}_{\ r}$, and $I^{\theta}_{\ \theta}$ 
with the ansatz for the physical and reference metrics as follows:
\begin{align}
\label{modification_t}
I^t{}_t \equiv& \beta_0 
+ \beta_1 \left( \frac{2\chi}{r} + \chi^{\prime} \e^{-\lambda}\right)
+ \beta_2 \left(\frac{\chi^2}{r^2}
+ \frac{2\chi\chi^{\prime}}{r}\e^{-\lambda}\right)
+ \beta_3 \frac{\chi^2\chi^{\prime}}{r^2}\e^{-\lambda}
\, , \\
\label{modification_r}
I^r{}_r \equiv& \beta_0
+ \beta_1\left( \frac{2\chi}{r} + \e^{-\nu}\right)
+ \beta_2\left( \frac{\chi^2}{r^2} + \frac{2\chi}{r}\e^{-\nu}\right)
+ \beta_3 \frac{\chi^2}{r^2}\e^{-\nu}
\, , \\
\label{modification_theta}
I^\theta{}_\theta 
\equiv& \beta_0+\beta_1\pqty{\frac{\chi}{r}+\chi^{\prime}\e^{-\lambda}+\e^{-\nu}}
+ \beta_2\pqty{\frac{1}{r}\chi\chi^{\prime}\e^{-\lambda}+\frac{1}{r}\chi \e^{-\nu}
+\chi^{\prime}\e^{-\lambda-\nu}}
+ \beta_3\frac{\chi\chi^{\prime}}{r}\e^{-\lambda-\nu} 
\, .
\end{align}

In addition to the equations of motion for the $tt$ and $rr$ components,
we need to take into account the divergence of equations of motion, 
\begin{equation}
\nabla_\mu\left(G^{\mu\nu}+m_0^2I^{\mu\nu}\right)=\kappa^2\nabla_\mu T^{\mu\nu}\, .
\end{equation}
If we assume the conservation of the energy-momentum tensor $\nabla_\mu T^{\mu\nu}=0$,
we obtain the new algebraic equations
\begin{equation}
\nabla_\mu I^{\mu\nu} = 0
\label{constraint}
\end{equation}
from the Bianchi identities $\nabla_\mu G^{\mu\nu}=0$.
Substituting Eqs.~\eqref{modification_t}, \eqref{modification_r}, and \eqref{modification_theta} 
into \eqref{constraint}, 
we find that $t$, $\theta$, and $\phi$ components of Eq.~\eqref{constraint} are identically satisfied,
and that the nontrivial $r$ component leads to the following equation:
\begin{align}
0=&
(\beta_1 r^2 + 2\beta_2 r\chi + \beta_3\chi^2) (\e^\nu)^{\prime} \nonumber \\
&
+ \left[ 2\beta_2(\e^\nu - \e^{\lambda + \nu}) + 2\beta_3(1-\e^\lambda) \right] \chi 
+ 2\beta_1 r (\e^\nu - \e^{\lambda+\nu}) + 2\beta_2 r (1-\e^\lambda) 
\, .
\label{constraint2}
\end{align}
The $\nu'$ contains the modification term $I^r{}_r$ as in Eq.~\eqref{eq:mtov_pres},
and it can be written by up to second-order non-derivative terms for $\chi$ as given 
in Eq.~\eqref{modification_r}.
Therefore, the new constraint displays the fourth-order algebraic equation for $\chi$.

Because the new constraint equation is the fourth-order algebraic equation,
we can solve it analytically.
For the convenience in the order estimation,
we replace the variables to dimensionless ones as follows:
\begin{align}
&r_g\equiv GM_\odot
\Leftrightarrow
\kappa^2=8\pi\frac{r_g}{M_\odot},\,
r\rightarrow rr_g,\,
\chi\rightarrow \chi r_g,\,
M(r)\rightarrow M(r)M_\odot, \nonumber \\
&\rho\rightarrow \rho\pqty{M_\odot/r_g^3},\,
p\rightarrow p\pqty{M_\odot/r_g^3},\,
m_0\rightarrow \frac{m_0}{r_g} \, .
\label{dimensionless}
\end{align}
Here, we note that the dimension of graviton mass is $\left[L^{-1}\right]$ because of our units, 
and the magnitude of the dimensionless graviton mass is very tiny such as
\begin{equation}
m_0
\sim 10^{-23}
\end{equation}
because we assume that the graviton mass is of order of the observed dark energy.
Since we also demand that $\beta_{n} = \order{1}$,
the modification of gravity seems to give the tiny effects to the observables.
However, the additional algebraic equation \eqref{constraint2},
which cannot be found in the general relativity, changes the mathematical structures 
of the equations of motion.

\section{Absence of the Vainshtein Mechanism in Minimal Model}

To solve Eqs.~\eqref{eq:mtov_mass}, \eqref{eq:mtov_pres}, and \eqref{constraint2} and 
obtain the mass-radius relation of the relativistic star, 
we have to construct the solutions with a specific equation of state numerically.
The typical way is imposing boundary condition and solving it as a two-point boundary 
value problem.
One of the two points in our case is the center of the relativistic star, and we impose 
Eq.~\eqref{boundary_condition1}.
Another is the point far away from the star (analytically, at infinity), 
thus, we need to check the asymptotic behaviors of the solutions.
If the Vainshtein mechanism works outside the star,
we can impose the boundary condition so that the Schwarzschild solution describes 
the space-time outside the star.

When we solve the equation of motion, 
the fact that the relativistic star system in massive gravity has two scales makes 
the analysis complicated.
These are the solar mass $M_\odot$, which characterize the astrophysical scale, 
and the graviton mass $m_0$, which characterize the cosmological scale.
In this section, instead of solving the system numerically, 
we evaluate the behavior of $\chi(r)$ near and far from the star by approximations analytically.
When we find the whole structure of the solution in the relativistic star system, 
it allows us to examine whether the screening mechanism can work or not,
which determines the appropriate boundary condition outside the relativistic stars 
in the dRGT massive gravity.

\subsection{$\chi(r)$ for Asymptotically Flat Space-time}
\label{sec_3A}

Before we study the asymptotic behavior of the space-time,
we think of a link between $g_{\mu \nu}$ and $\chi(r)$.
First, when we make an assumption 
that $\e^{2\nu}=\e^{2\lambda}=1$ in the physical metric $g_{\mu \nu}$ and $\chi(r) = r$ 
in the reference metric $f_{\mu \nu}$,
one finds that it is consistent with the equation of motion
because $g_{\mu \nu} = f_{\mu \nu} = \eta_{\mu \nu}$ is the solution with generic choices 
of the parameters $\beta_{n}$~\cite{Katsuragawa:2013lfa}.
In the above case, one finds
\begin{equation}
I_{\mu \nu} = (\beta_{0} + 3 \beta_{1} + 3\beta_{2} + \beta_{3} ) E_{(4) \mu \nu} \, ,
\end{equation}
where $E_{(4)}$ represents the $4 \times 4$ identity matrix. 
Eq.~\eqref{beta2} leads
\begin{equation}
\beta_{0} + 3 \beta_{1} + 3\beta_{2} + \beta_{3} = 0 \, ,
\end{equation}
where the modification terms vanish.
Because the energy-momentum tensor $T_{\mu \nu}$ also vanish outside the star, 
the equation of motion \eqref{eom} reduces to the Einstein equation in the vacuum.

However, it is not trivial that the asymptotic flatness in $g_{\mu \nu}$ is identical to the condition 
$\chi(r) = r$.
Next, we only assume that the physical space-time shows the asymptotic flatness outside 
the star, $g_{\mu \nu}=\eta_{\mu \nu}$.
Because the Einstein tensor $G_{\mu \nu}$ and energy-momentum tensor $T_{\mu \nu}$ vanish,
the modification terms should vanish, $I^{\mu \nu} = 0$, which suggests that 
Eq.~\eqref{constraint} also satisfies.
Thus, we find Eqs.~\eqref{modification_t},\eqref{modification_r}, and \eqref{modification_theta} lead
\begin{align}
0=& \beta_0 
  + \beta_1 \left( \frac{2\chi}{r} + \chi^{\prime} \right) 
  + \beta_2 \left( \frac{\chi^2}{r^2}  + \frac{2\chi\chi^{\prime}}{r} \right)
  + \beta_3 \frac{\chi^2\chi^{\prime}}{r^2}
\label{modification_t2}
 \, , \\
0
=& \left( \beta_0 + \beta_1 \right) + 2\left(\beta_1 + \beta_2 \right) \frac{\chi}{r} 
+ \left( \beta_2 + \beta_3 \right) \left( \frac{\chi}{r} \right)^2 
\label{modification_r2}
\, , \\
0=& \left( \beta_0 + \beta_1 \right)
+ \left(\beta_1 + \beta_2 \right)  \left( \frac{\chi}{r}+\chi^{\prime} \right)
+  \left( \beta_2 + \beta_3 \right) \frac{\chi\chi^{\prime}}{r}
\label{modification_theta2}
\, .
\end{align}
Using Eq.~\eqref{beta2} for Eq.~\eqref{modification_r2}, we obtain
\begin{align}
0
=& \left(\frac{\chi}{r} - 1 \right) \left[ \left( \beta_0 + \beta_1 - 2 \right) 
\frac{\chi}{r} - \left( \beta_0 + \beta_1\right) \right] 
\end{align}
and the solutions are
\begin{equation}
\frac{\chi}{r} = 1\, , \quad  \frac{\beta_0 + \beta_1}{\beta_0+\beta_1-2} \, .
\label{chi_r}
\end{equation}
Note that we have only the first solution in the minimal model
because the second one diverges.
Moreover, we always have $\chi/r=1$ in the case $\beta_0+\beta_1=2$.
The second solution does not give $\chi/r=1$ when $\beta_0+\beta_1$ takes finite value.
In any models, one can find that the $\chi$ should take the form of $\chi(r) = A r$, 
where $A$ is a constant.

When we substitute this linear solution to Eq.~\eqref{modification_theta2}, we find
\begin{equation}
0
=\left(A - 1 \right) \left[ \left( \beta_0 + \beta_1 - 2 \right) A 
 - \left( \beta_0 + \beta_1\right) \right] 
\, ,
\end{equation}
and thus, we obtain the results identical with Eq.~\eqref{chi_r},
\begin{equation}
A=1\, , \quad \frac{\beta_0 + \beta_1}{\beta_0+\beta_1-2} \, .
\end{equation}
By substituting $\chi(r) = A r$ to Eq.~\eqref{modification_t2}, 
we find
\begin{align}
0
=& \beta_0 + 3\beta_1 A + 3 \beta_2 A^2 + \beta_3 A^3 
\nonumber \\
=& (A-1) \left[  \left(2\beta_0+3\beta_1-3\right) A^2 - \left( \beta_{0} 
+ 3\beta_{1} \right) A- \beta_0 \right]
\, .
\end{align}
One solution is $A=1$ and the others satisfy the following equation, 
\begin{align}
\left(2\beta_0+3\beta_1-3\right) A^2 - \left( \beta_{0} + 3\beta_{1} \right) A- \beta_0 = 0 \, .
\label{modification_t2-2}
\end{align}
Note that $A=1$ does not satisfy Eq.~\eqref{modification_t2-2} in any choice of 
$\beta_{0}$ and $\beta_{1}$.
When we substitute the second solution of Eqs.~\eqref{modification_r2} 
or \eqref{modification_theta2} to Eq.~\eqref{modification_t2-2},
we obtain the consistent solution $\chi(r) = A r$ for the specific choice of 
$\beta_{0}$ and $\beta_{1}$.

We have found that $\chi(r) = r$ with the generic parameters and that $\chi(r)=Ar$ 
with $A \neq 1$ for the specific parameters
when we require the asymptotic flatness for $g_{\mu \nu}$.
If we substitute $\chi(r) = Ar$ in Eq.~\eqref{reference_metric}, 
the reference metric takes the following form
\begin{align}
f_{\mu\nu} d  x^\mu  d  x^\nu
&= - d  t^2 + A^2 d  r^2 + A^2r^2 d \Omega^2 \, ,
\end{align}
If one redefines the radial coordinate $Ar \rightarrow r$, we can remove the factor $A$ 
in the reference metric.
We can absorb the scaling by $A$  into the scaling ambiguity of the definition of $\chi(r)$,
therefore, $\chi(r)=Ar$ also represents the Minkowski space-time in the reference metric 
$f_{\mu \nu}$.
We also note that the scaling factor $A$, which is determined by $\beta_n$, is of order of unity 
when we assume $\beta_n = \order{1}$ and do not use the specified choice 
so that $\beta_{0} + \beta_{1} \approx 2$.
In the following, we calculate the case of $A=1,\, \chi(r)=r$ for simplicity.

Finally, we consider the inverse problem and only assume that $\chi(r) = r$ outside the star.
Because $p$ and $M^{\prime}$ vanishes,
the equations of motion Eqs.~\eqref{eq:mtov_mass}, \eqref{eq:mtov_pres}, 
and \eqref{eq:mtov_theta} give
\begin{align}
0&= \beta_0 
+ \beta_1 \left( 2 + \e^{-\lambda}\right)
+ \beta_2 \left( 1 + 2\e^{-\lambda}\right)
+ \beta_3 \e^{-\lambda}
\label{eq:mtov_mass2}
\, , \\
2r \e^{2\lambda} \nu' 
&=  \left( 1 - \e^{-2\lambda} \right) 
- m_0^2r^2
\left[ \beta_0
+ \beta_1 \left( 2 + \e^{-\nu}\right)
+ \beta_2 \left( 1 + 2 \e^{-\nu}\right)
+ \beta_3 \e^{-\nu}
 \right]
\label{eq:mtov_pres2}
\, , \\
0&=\left( \nu''+\nu'^2+\frac{\nu'}{r} \right) \e^{-2\lambda}
+ \frac{1}{2}\pqty{\frac{1}{r}+\nu'}(\e^{-2\lambda})' \nonumber \\
& \quad +m_0^2
\left[ 
\beta_0
+ \beta_1 \left( 1 + \e^{-\lambda} + \e^{-\nu} \right)
+ \beta_2 \left( \e^{-\lambda} + \e^{-\nu} + \e^{-\lambda-\nu} \right)
+ \beta_3 \e^{-\lambda-\nu} 
\right] 
\, .
\label{eq:mtov_theta2}
\end{align}
Here, we have used $\e^{-2\lambda} \equiv 1-2GM/r$.
When we substitute Eq.~\eqref{beta2} into Eq.~\eqref{eq:mtov_mass2},
we find
\begin{equation}
\e^{-\lambda}= 1 \, , 
\label{eq:mtov_mass3}
\end{equation}
furthermore, Eqs.~\eqref{eq:mtov_pres2} and \eqref{eq:mtov_pres2} are given by
\begin{align}
0&= \frac{2\nu'}{r} + m_0^2 \left( 1 - \e^{-\nu} \right)
\label{eq:mtov_pres3}
\, , 
\\
0&= \left( \nu''+\nu'^2+\frac{\nu'}{r} \right)
+ m_0^2 \left( 1 - \e^{-\nu} \right)
\, .
\label{eq:mtov_theta3}
\end{align}
For the general $m_{0}^{2}$, we only find the trivial solution:
\begin{equation}
\e^{\nu}=1 \, .
\end{equation}
Based on the discussion in this subsection, 
we have found that asymptotic flatness in $g_{\mu \nu}$ is equivalent to $\chi(r)=r$,
which allows us to study the behavior of the physical space-time in terms of the $\chi(r)$.
If $\chi(r)$ shows the asymptotically flat feature,
we can infer and conclude that the physical space-time is also the asymptotically flat.

\subsection{Asymptotic Behavior Near and Away from Stars}
\label{sec_3B}

To study the asymptotic behavior of $\chi(r)$, 
we introduce a mass scale $M_{s}$, to denote the dimensionless mass of the relativistic star.
Since we formulated the equations in terms of dimensionless quantities normalized 
by the solar-scale,
and we expect $M_{s}$ is also at the solar-scale scale, $M_{s} = \order{1}$.
Around the object with a particular mass scale, 
we can introduce the significant scale, the Vainshtein radius.
In the dRGT massive gravity, the Vainshtein radius is defined by
\begin{equation}
r_{V} = \left( \frac{M_{s}}{M_\mathrm{Pl}} \right)^{1/3} \frac{1}{\Lambda_{3}}
\, .
\label{Vainshtein_radius}
\end{equation}
where$M_\mathrm{Pl}$ is the Planck mass and 
$\Lambda_{3}$ is the cutoff scale in the dRGT massive gravity, defined as
\begin{equation}
\Lambda_{3} = \left( M_\mathrm{Pl} m_{0}^{2} \right)^{1/3}
\, .
\label{cutoff}
\end{equation}
In our normalization, rescaled by the solar-mass scale, we find
\begin{equation}
r_{V}= \left( \frac{M_{s}}{m_{0}^{2}} \right)^{1/3} \, .
\end{equation}
Assuming $M_{s} = \order{1}$, the Vainshtein radius is $r_{V} = m_{0}^{-2/3} \sim 10^{15}$

As we have mentioned, 
the possible difficulty is that the Vainshtein radius is the product of astrophysical $M_{s}$ 
and cosmological scales $m_{0}$.
In order to deal with the important intermediate scale $r_{V}$, 
we focus on the scale $M_{s} \ll r \ll r_{V}$, 
to address the space-time outside but not far away from the star.
If the Vainshtein mechanism works, the dRGT massive gravity restores the results 
in the general relativity inside the Vainshtein radius, and the physical space-time should be 
the Schwarzschild space-time.
Thus, in the region $M_{s} \ll r \ll r_{V}$,
we assume that the physical metric is given by the the Schwarzschild space-time, 
\begin{equation}
\label{eq:asymp_nu_and_lambda}
\e^{2\nu(r)} = 1 - \frac{2M_{s}}{r} 
\, , \quad 
\e^{-2\lambda(r)}= 1 - \frac{2M_{s}}{r} 
\, .
\end{equation}
$M_{s}/r \ll 1$ in $M_{s} \ll r \ll r_{V}$, 
and we can treat $M_{s}/r$ as the perturbation from the Minkowski space-time.

Furthermore, the discussion in the previous subsection implies $\chi(r)$ should take 
the following form: 
\begin{equation}
\frac{\chi(r)}{r} = 1 + \order{\frac{M_{s}}{r}}
\label{eq:expansion0}
\, .
\end{equation}
to balance the order of the perturbations in both-hand sides of the equations of motion.
Note that we can rescale $\chi(r)$ with the arbitrary factor to express the above form 
if it is necessary.
In other words,
when we find the above $\chi(r)$ as a solution to the equation of motion in $M_{s} \ll r \ll r_{V}$,
we have the Schwarzschild space-time outside the star, 
which suggests the Vainshtein mechanism works properly.
If the $\chi(r)$ shows the large deviation from the asymptotic form $\chi(r) = r$,
it implies that the Vainshtein mechanism does not work.

\subsection{Asymptotic Flatness Around and Away from the Stars}
\label{sec_3C}

Before we discuss the relativistic star in the general case of the dRGT massive gravity,
we consider the minimal model in which the parameters $\beta_{n}$ are chosen as in 
Eq.~\eqref{minimal_model}.
In our previous work, 
we directly derived the mass-radius relation in the minimal model by the numerical simulation 
and discussed the effect of the modification on the maximal mass of the stars.
Here, we refine our previous result from the viewpoints of the boundary conditions and 
the Vainshtein mechanism.

Substituting $\beta_{n}$ in the minimal model \eqref{minimal_model} into the Eq.~\eqref{constraint2},
we obtain the following algebraic equation,
\begin{equation}
\nu^{\prime} = \frac{2}{r}\pqty{\e^\lambda-1}  \, .
\end{equation}
And, Eq.~\eqref{eq:mtov_pres} is given by
\begin{equation}
\nu'
=\frac{1}{2}\pqty{\kappa^2pr+\frac{1}{r}-m_0^2rI^r{}_r}\e^{2\lambda}-\frac{1}{2r}\, ,\quad
I^r{}_r=3-\pqty{\frac{2\chi}{r}+\e^{-\nu}}\, .
\end{equation}
By eliminating $\nu^{\prime}$, we find the first-order algebraic equation for $\chi$,
whose solution is
\begin{equation}
\chi=
\frac{r}{2} \left \{ 3-\e^{-\nu}+\frac{1}{m_0^2r^2} \left[ (4\e^\lambda-3)
\e^{-2\lambda}-1-\kappa^2pr^2 \right] \right \} 
\label{eq:chi_sol_minimal}
\, .
\end{equation}
This result shows that the minimal model does not have the Vainshtein mechanism
because the additional terms proportional to $1/(m_{0}^{2}r^{2})$ becomes relevant at 
the small scale $r \ll 1/m_{0}$, 
including the interior region of the star.

As an illustration, we assume the Schwarzschild space-time outside the star.
If this assumption is appropriate, we get the Eq.~\eqref{eq:expansion0} from 
Eq.~\eqref{eq:chi_sol_minimal}.
Substituting Eq.~\eqref{eq:asymp_nu_and_lambda} and $p=0$, we obtain 
\begin{align}
\frac{\chi}{r} 
=&
1 - \frac{M_{s}}{2r} + \order{\frac{M_{s}^2}{r^2}}
+ \frac{1}{m_{0}^{2}r^{2}} \left[ \frac{M_{s}}{r} + \order{\frac{M_{s}^2}{r^2}} \right] 
\nonumber \\
=&
\left( \frac{r_{V}}{r} \right)^3 \left[ 1 + \order{\frac{M_{s}}{r}} \right] + 1 - \frac{M_{s}}{2r} 
+ \order{\frac{M_{s}^2}{r^2}} 
\label{minimal_chi}
\, .
\end{align}
Eq.~\eqref{minimal_chi} displays the significant deviations from Eq.~\eqref{eq:expansion0} 
because the first term becomes dominant inside the Vainshtein radius, $r_{V}/r \gg 1$.
It suggests that the physical metric $g_{\mu \nu}$ does not describe the Schwarzschild space-time
outside the star.
Therefore, we can conclude that the Vainshtein mechanism does not work in the minimal model 
with the flat reference metric.

\section{Screened and Unscreened Solutions in Non-Minimal Model}

In the previous section,
we have discussed the physical space-time around the star in the context of $\chi(r)$.
We have found that the Vainshtein mechanism does always not work around the star 
in the minimal model, where $\beta_{n}$'s are specially chosen.
In this section, we consider the general case, the non-minimal model of the dRGT massive gravity 
and check the asymptotic behavior of $\chi(r)$ and examine the Vainshtein mechanism.

\subsection{Fourth-Order Equation for $\chi(r)$}
\label{sec_4A}

To solve the new algebraic equation, 
we eliminate the $\nu^{\prime}$ from Eq.~\eqref{eq:mtov_pres} with Eq.~\eqref{constraint2}
as we performed in the case of the minimal model.
For the convention, we express Eq.~\eqref{eq:mtov_pres} in the following form written 
in the new variables,
\begin{equation}
\nu' \equiv n_0 -m_0^2 n_1 I^r{}_r \quad \left(n_0,\,n_1>0 \right)
\, ,
\end{equation}
where we define
\begin{equation}
n_0 \equiv \frac{1}{2} \kappa^2 pr\e^{2\lambda} + \frac{1}{2r} \left(\e^{2\lambda} - 1\right)
\, , \quad 
n_1 \equiv \frac{1}{2}r\e^{2\lambda}
\, .
\end{equation}
Using Eqs.~\eqref{eq:mtov_pres} and \eqref{modification_r}, 
we find that Eq.~\eqref{constraint2} in the generic case of the parameters $\beta_{n}$ leads to
\begin{align}
0
&=
(\beta_1 r^2 + 2\beta_2 r\chi + \beta_3\chi^2) \e^{\nu} \left( n_0 -m_0^2 n_1 I^r{}_r \right)
\nonumber \\
& 
+ \left[ 2\beta_2(\e^\nu - \e^{\lambda + \nu}) + 2\beta_3(1-\e^\lambda) \right] \chi 
+ 2\beta_1 r (\e^\nu - \e^{\lambda+\nu}) + 2\beta_2 r (1-\e^\lambda) 
\nonumber \\
&=
 -m_0^2n_1\e^\nu (\beta_3\chi^2+2\beta_2r\chi+\beta_1r^2) \nonumber\\
& 
\times \left[ \frac{1}{r^2}(\beta_2+\beta_3\e^{-\nu})\chi^2
+ \frac{2}{r}(\beta_1+\beta_2\e^{-\nu})\chi + (\beta_0+\beta_1\e^{-\nu}) \right] 
\nonumber \\
& 
+ \beta_3 n_0\e^\nu\chi^2 + \left[ 2\beta_2r n_0 \e^\nu 
+ 2\beta_2(\e^\nu - \e^{\lambda + \nu}) + 2\beta_3(1-\e^\lambda) \right] \chi 
\nonumber \\
& 
+ 2\beta_1 r (\e^\nu - \e^{\lambda+\nu}) + 2\beta_2 r (1-\e^\lambda)  + \beta_1 r^2 n_0 \e^\nu 
\, .
\label{algeb}
\end{align}
Expanding the above expression as the polynomial with respect to $\chi$, 
we obtain the following the fourth-order algebraic equation,
\begin{align}
0=& 
- \frac{m_0^2 n_1 \e^{\nu}}{r^2} \beta_3(\beta_2+\beta_3\e^{-\nu})\chi^4
- \frac{2m_0^2 n_1 \e^{\nu}}{r} \left[ 
\beta_2 (\beta_2+\beta_3\e^{-\nu}) + \beta_3(\beta_1+\beta_2\e^{-\nu}) 
\right] \chi^{3}
\nonumber \\
&
+ \left \{
\beta_3 n_0 \e^{\nu} 
- m_0^2 n_1 \e^{\nu} \left[
\beta_1 (\beta_2+\beta_3\e^{-\nu}) 
- 4 \beta_2 (\beta_1+\beta_2\e^{-\nu}) 
- \beta_3\left(\beta_0+\beta_1\e^{-\nu} \right)
\right]
\right \} \chi^2
\nonumber \\
&
+ \left \{
2 \left[ \beta_2 n_0 r \e^{\nu} + \beta_2(\e^\nu - \e^{\lambda + \nu}) 
+ \beta_3(1-\e^\lambda) \right] 
\right.
\nonumber \\
& \left. - 2m_0^2 n_1r \e^{\nu} \left[
\beta_1 (\beta_1+\beta_2\e^{-\nu}) 
+ \beta_2\left(\beta_0+\beta_1\e^{-\nu} \right)
\right] 
\right \} \chi 
\nonumber \\
&
+ r \left[
2\beta_1 (\e^\nu - \e^{\lambda+\nu}) + 2\beta_2 (1-\e^\lambda) + \beta_1 n_0 r \e^{\nu} 
\right]
 -m_0^2 n_1 r^2 \e^{\nu} \beta_1\left(\beta_0+\beta_1\e^{-\nu} \right) \, .
\label{algeb2}
\end{align}
As we mentioned below Eq.~(\ref{minimal_model}),
we need to choose $\beta_{2}, \beta_3 \neq 0$ to realize the non-minimal model of 
the dRGT massive gravity.
For this restriction of $\beta_{2}$ and $\beta_{3}$, 
we find that Eq.~\eqref{algeb2} is the fourth order with respect to $\chi$.
Note that in the minimal model, 
one can confirm that Eq.~\eqref{algeb2} is indeed reduced to the first-order equation, 
which restores Eq.~\eqref{eq:chi_sol_minimal}.

For the further convenience in the later calculation,
we rewrite Eq.~\eqref{algeb2} with normalizing the coefficient of the $\chi^4$ term:
\begin{equation}
\chi^4
+a\chi^3
 -\frac{1}{m_0^2}\bqty{\pqty{b_0 + m_0^2b_1}\chi^2
+\pqty{c_0 + m_0^2c_1}\chi
+\pqty{d_0 + m_0^2d_1}}=0 \, ,
\label{chi_eq}
\end{equation}
where we define the coefficients as follows,
\begin{align}
a&=\frac{ 2r \left[ \beta_3(\beta_1+\beta_2\e^{-\nu})+\beta_2(\beta_2
+\beta_3\e^{-\nu}) \right]}{\beta_3(\beta_2+\beta_3\e^{-\nu})} 
\, , \nonumber \\
b_0&= \frac{n_{0} r^2}{n_{1} \left(\beta_2+\beta_{3}\e^{-\nu} \right)}
\, , \nonumber \\
b_{1}&= - \frac{ r^{2} \left[ 4 \beta_{2}^{2}\e^{-\nu} + \beta_{0} \beta_{3} + \beta_{1} (5 \beta_{2} 
+ 2 \beta_{3} \e^{-\nu}) \right]}
{\beta_{3} (\beta_{2} + \beta_{3}\e^{-\nu} )} 
\, , \nonumber \\
c_0&=\frac{2 r^2 [\beta_2r n_0  + \beta_2(1 - \e^{\lambda}) 
+ \beta_3(\e^{-\nu}-\e^{\lambda - \nu})] }
{n_{1} \beta_3(\beta_2+\beta_3\e^{-\nu})}
\, , \nonumber \\
c_{1}&= - \frac{2 r^{3} \left[\beta_{2} (\beta_{0} + \beta_{1} \e^{-\nu}) + \beta_{1} (\beta_{1}
+  \beta_{2} \e^{-\nu}) \right] }
{\beta_{3} (\beta_{2} + \beta_{3}\e^{-\nu} )} 
\, , \nonumber \\
d_0&=\frac{r^3[\beta_1 r n_0 + 2\beta_1 \left(1- \e^{\lambda} \right) 
+ 2 \beta_2 (\e^{- \nu} - \e^{\lambda- \nu}) ]}
{n_1 \beta_3(\beta_2+\beta_3\e^{-\nu})}
\, , \nonumber \\
d_{1}&= - \frac{r^{4} \beta_{1} (\beta_{0} + \beta_{1} \e^{-\nu} )}{\beta_{3} (\beta_{2} 
+ \beta_{3}\e^{-\nu} )}
\, .
\label{coefficients1}
\end{align}
When we obtain the real solutions of Eq.~\eqref{chi_eq} and study their asymptotic behavior 
away from the star,
we can discuss the physical spce-time to connect the Schwarzschild space-time
as we have done in the case of the minimal model of the dRGT massive gravity.

\subsection{Brunch Analysis for $\chi$ Around Star}
\label{sec_4B}

As we have performed in the minimal model, 
we assume the Schwarzschild space-time in the region $M_{s} \ll r \ll r_{V}$ 
and study $\chi(r)$ outside the star in the non-minimal model.
If $\chi(r)$ shows the asymptotic behavior as expected in Eq.~\eqref{eq:expansion0},
we can conclude that the Vainshtein mechanism works in the non-minimal model, 
otherwise, the screening mechanism does not work in the general model 
of the dRGT massive gravity.

Compared with the minimal model,
we have a remarkable difficulty to obtain $\chi(r)$ in the non-minimal model
because of the higher-order algebraic equation Eq.~\eqref{chi_eq}.
To make it manageable in an analytical manner, 
we begin the analysis with the assumption that the physical space-time is described 
by the Schwarzschild solution outside the star,
instead of looking for the exact solutions.
Then, we examine the asymptotic behavior of $\chi(r)$ and check whether it is consistent 
with the assumption.
When we use Eq.~\eqref{eq:asymp_nu_and_lambda} with the condition $p=0$, 
we find
\begin{equation}
n_{0} = \frac{1}{r} \left[ \frac{M_{s}}{r} + \order{\frac{M_{s}^{2}}{r^{2}}} \right] 
\, , \quad 
\frac{1}{n_{1}} = \frac{2}{r} \left[ 1 - \frac{2M_{s}}{r} \right]
\, .
\end{equation}
Furthermore, when we use $\beta_{n} = \order{1}$, 
we can expand the coefficients of the fourth-order equation Eq.~\eqref{coefficients1} in terms 
of $\order{M_{s}/r}$ as follows, 
\begin{align}
\label{coefficients2}
a&= 2r \left[ A  + \tilde{A} \frac{M_{s}}{r} + \order{\frac{M_{s}^2}{r^2}}  \right]
\, , \nonumber \\
b_0& = B_0 \frac{M_{s}}{r} + \order{\frac{M_{s}^2}{r^2}} 
\, , \quad  
b_{1}= - r^{2} \left[ B_{1} + \tilde{B}_{1} \frac{M_{s}}{r} + \order{\frac{M_{s}^2}{r^2}} \right]
\, , \nonumber \\
c_0&= r C_{0} \left[ \frac{M_{s}}{r} + \order{\frac{M_{s}^2}{r^2}}  \right]
\, , \quad 
c_{1}= - 2 r^{3} \left[ C_{1} + \tilde{C}_{1} \frac{M_{s}}{r} + \order{\frac{M_{s}^2}{r^2}} \right]
\, , \nonumber \\
d_0&= r^2 D_{0} \left[ \frac{M_{s}}{r} + \order{\frac{M_{s}^2}{r^2}}  \right]
\, , \quad 
d_{1}= - r^{4} 
\left[
D_{1} 
+ \tilde{D}_{1} \frac{M_{s}}{r} 
+ \order{\frac{M_{s}^2}{r^2}}
\right]
\, ,
\end{align}
where
\begin{align}
\label{coefficients3}
A&=
\frac{ \beta_3(\beta_1+\beta_2)+\beta_2(\beta_2+\beta_3) }{\beta_3(\beta_2+\beta_3)} 
\, , \quad 
\tilde{A}=
\frac{2 \beta_2 \beta_3 - \beta_3^2 A  }{\beta_3\left( \beta_2 + \beta_3 \right) }
\, , \nonumber \\
B_{0}&= 
\frac{2\beta_3}{\beta_3 \left( \beta_2 + \beta_3 \right) }
\, , \quad 
B_{1} = 
\frac{4 \beta_{2}^{2} + \beta_{0} \beta_{3} + \beta_{1} (5 \beta_{2} + 2 \beta_{3} ) }
{\beta_{3} (\beta_{2} + \beta_{3})} 
\, , \quad 
\tilde{B}_{1}= 
\frac{\left(  4\beta_{2}^{2} + 2 \beta_{1}\beta_{3} \right) - \beta_3^2 B_{1}}{\beta_{3}\left( \beta_2 
+ \beta_3 \right) }
\, , \nonumber \\
C_{0}&= 
\frac{-4 \beta_3}{\beta_3\left( \beta_2 + \beta_3 \right)}
\, , \quad 
C_{1} = 
\frac{ \beta_{2} (\beta_{0} + \beta_{1} ) + \beta_{1} (\beta_{1}+  \beta_{2})}
{\beta_{3} (\beta_{2} + \beta_{3})} 
\, , \quad 
\tilde{C}_{1}= 
\frac{2 \beta_{1} \beta_{2} - \beta_3^2C_{1}}{\beta_{3}\left( \beta_2 + \beta_3 \right)}
\, , \nonumber \\
D_{0}&= 
\frac{- 2\beta_1 - 4 \beta_2}{\beta_{3}\left( \beta_2 + \beta_3 \right)} 
\, , \quad 
D_{1}= 
\frac{ \beta_{1} (\beta_{0} + \beta_{1} )}{\beta_{3} (\beta_{2} + \beta_{3} )}
\, , \quad 
\tilde{D}_{1}= 
\frac{\beta_{1}^2 - \beta_3^2 D_{1}}{\beta_{3}\left( \beta_2 + \beta_3 \right)} 
\, .
\end{align}
Therefore, when we assume the Schwarzschild space-time for the physical metric $g_{\mu \nu}$, 
the fourth-order equation takes the following form:
\begin{align}
\label{chi_eq2}
0=&
\left( \frac{\chi}{r}\right)^4
+ \left[ 2 A + 2 \tilde{A} \frac{M_{s}}{r} + \order{\frac{M_{s}^2}{r^2}} \right] 
\left( \frac{\chi}{r}\right)^3
\nonumber \\
& 
 - \frac{1}{m_{0}^{2}r^{2}} 
\left\{
\left[B_0 \frac{M_{s}}{r} + \order{\frac{M_{s}^2}{r^2}}  \right] 
 - m_{0}^{2}r^{2} \left[ B_{1} + \tilde{B}_{1} \frac{M_{s}}{r} + \order{\frac{M_{s}^2}{r^2}} \right]
\right\}
\left( \frac{\chi}{r}\right)^2
\nonumber \\
&
 - \frac{1}{m_{0}^{2}r^{2}} 
\left \{
\left[C_{0}  \frac{M_{s}}{r} + \order{\frac{M_{s}^2}{r^2}}  \right] 
 -m_{0}^{2}r^{2} \left[ 2 C_{1} + 2\tilde{C}_{1} \frac{M_{s}}{r} + \order{\frac{M_{s}^2}{r^2}} \right]
\right \}
\left( \frac{\chi}{r}\right)
\nonumber \\
& 
 - \frac{1}{m_{0}^{2}r^{2}}
\left \{
\left[D_{0}  \frac{M_{s}}{r} + \order{\frac{M_{s}^2}{r^2}}  \right] 
 - m_{0}^{2}r^{2} \left[ D_{1} + \tilde{D}_{1} \frac{M_{s}}{r} + \order{\frac{M_{s}^2}{r^2}} \right]
\right \}
\, .
\end{align}
We find that $1/(m_{0}^{2} r^{2})$ corrections show up as in Eq.~\eqref{eq:chi_sol_minimal} 
for the minimal model,
which would bring the origin of the large deviation from the asymptotic flatness.
Noting $m_{0}^{2} r^{2}  \ll  M_{s}/r \ll 1$ in the region $M_{s} \ll r \ll r_{V}$
because
\begin{equation}
\frac{M_{s}/r}{m_{0}^{2} r^{2}} = \left( \frac{r_{V}}{r} \right)^{3} \gg 1
\, ,
\end{equation}
the fourth-order equation Eq.~\eqref{chi_eq2} can be further approximated and given by
\begin{align}
\label{chi_eq3}
0=&
\left( \frac{\chi}{r}\right)^4
+ \left[ 2 A + 2 \tilde{A} \frac{M_{s}}{r} + \order{\frac{M_{s}^2}{r^2}} \right] 
\left( \frac{\chi}{r}\right)^3
\nonumber \\
&
- \left( \frac{r_{V}}{r} \right)^{3} \left \{ \left[B_0 
+ \order{\frac{M_{s}}{r}} \right] \left( \frac{\chi}{r}\right)^{2}
+ \left[C_{0} + \order{\frac{M_{s}}{r}} \right] \left( \frac{\chi}{r}\right)
+ \left[D_{0} + \order{\frac{M_{s}}{r}} \right] \right \} 
\, .
\end{align}

Finally, we solve the above algebraic equation Eq.~\eqref{chi_eq3} to $\chi(r)$.
If we assume the asymptotically flat solution for the reference metric $f_{\mu \nu}$, 
$\chi/r = \order{1} + \order{M_{s}/r}$ up to the scaling,
the first line of Eq.~\eqref{chi_eq3} is of order of $\order{1}$,
while the second line is order of $\order{(r_{V}/r)^{3}} \gg 1$.
Thus, the first line is negligible and the second line is dominant and we find
\begin{equation}
\label{chi_eq4}
0 =
\left[B_0 + \order{\frac{M_{s}}{r}}  \right] \left( \frac{\chi}{r}\right)^2
+ \left[C_{0} + \order{\frac{M_{s}}{r}}  \right] \left( \frac{\chi}{r}\right)
+ \left[D_{0}  + \order{\frac{M_{s}}{r}}  \right] 
\, ,
\end{equation}
and the solution is given by
\begin{equation}
\frac{\chi}{r} 
=
\frac{ - C_{0} \pm \sqrt{ C_{0}^2 - 4 B_0 D_{0} } }{ 2 B_{0} }  + \order{ \frac{M_{s}}{r} }
\, .
\end{equation}
Actually, the above solution is consistent with the assumption $\chi/r = \order{1} + \order{M_{s}/r}$,
and after rescaling the solution, we find
\begin{equation}
\label{nonminimal_solution1}
\frac{\chi}{r} = 1 + \order{\frac{M_{s}}{r}}
\, .
\end{equation}
This solution suggests that the non-minimal model of the dRGT massive gravity possesses 
the Vainshtein mechanism
around the relativistic star.
Therefore, one can study the relativistic star with the particular equation of state with the boundary 
condition to connect to the Schwarzschild space-time outside the star, as in the general relativity.
We emphasize that the existence of the real solutions depends on the parameters $\beta_{n}$'s 
whose region is evaluated with the condition that the determinant $D=C_{0}^2 - 4 B_0 D_{0} \geq 0$.
Moreover, we need to require that, at least, one of the two solutions is positive definite to express 
the radial coordinate.

On the other hand, in general, we have four solutions for Eq.~\eqref{chi_eq3}.
Since we have the two of the four, which are of the order of unity at the leading order,
we can analyze the leading order of the other two according to the coefficients of 
Eq.~\eqref{chi_eq3}.
When we express the four solutions as $\alpha$, $\beta$, $\gamma$, and $\delta$,
they satisfies,
\begin{align}
\alpha + \beta + \gamma + \delta 
&= - \left[ 2 A + \order{\frac{M_{s}}{r}}\right]
\, , \\
\alpha \beta + \alpha \gamma + \alpha \delta + \beta \gamma + \beta \delta + \gamma \delta
&= - \left( \frac{r_{V}}{r} \right)^{3}  \left[B_0 + \order{\frac{M_{s}}{r}} \right] 
\, , \\
\alpha \beta \gamma + \alpha \beta \delta + \alpha \gamma \delta + \beta \gamma \delta 
&= \left( \frac{r_{V}}{r} \right)^{3} \left[C_0 + \order{\frac{M_{s}}{r}} \right]
\, , \\
\alpha \beta \gamma \delta
&= - \left( \frac{r_{V}}{r} \right)^{3} \left[D_{0} + \order{\frac{M_{s}}{r}} \right] 
\, .
\end{align}
Furthermore, if we assume $\alpha$ and $\beta$ approximately obey Eq.~\eqref{chi_eq4},
we find
\begin{equation}
\alpha + \beta 
= - \frac{C_{0}}{B_{0}} + \order{\frac{M_{s}}{r}} 
\, , \quad 
\alpha \beta
= \frac{D_{0}}{B_{0}} + \order{\frac{M_{s}}{r}} 
\, .
\end{equation}
Thus, the sum and product of the other two solutions are given by
\begin{equation}
\gamma + \delta 
= \left( \frac{C_{0}}{B_{0}} - 2 A \right) + \order{\frac{M_{s}}{r}} 
\, , \quad 
\gamma \delta
= - \left( \frac{r_{V}}{r} \right)^{3} \frac{B_{0}}{D_{0}}
\, .
\end{equation}
In order to satisfy the above relation,
we can deduce the relevant expressions of the two solutions, $\gamma$ and $\delta$, as follows:
\begin{align}
\label{nonminimal_solution2}
\gamma &= \sqrt{\frac{B_{0}}{D_{0}}} \left( \frac{r_{V}}{r} \right)^{3/2}  + \order{1}
\, , \\
\delta &= - \sqrt{\frac{B_{0}}{D_{0}}} R \left( \frac{r_{V}}{r} \right)^{3/2}  + \order{1}
\, .
\end{align}
We find that these two solutions include the significant deviation 
from the asymptotically flat reference metric.
As in the minimal model,
we can understand that the non-minimal model includes the asymptotically non-flat solutions 
Eq.~\eqref{nonminimal_solution2} 
although it potentially possesses the asymptotically flat solutions Eq.~\eqref{nonminimal_solution1}.
We note again that the existence of the real solutions depends on the parameter choice $\beta_{n}$;
for instance, we would find the proper parameter regions so that $B_{0} / D_{0} >0$.

 From Eq.~\eqref{chi_eq3}, we have found the two different brunches of solutions for 
Eq.~\eqref{chi_eq3}, 
\begin{align}
\chi= \left( \frac{r_{V}}{r} \right)^{3/2} +\order{1} \, , \ 1 + \order{\frac{M_{s}}{r}}
\, .
\label{chi_solution}
\end{align}
The former has the large correction $\order{(r_{V}/r)^{3/2} } \gg 1$ in the region 
of our interest $M_{s} \ll r \ll r_{V}$
although the latter is of order of unity.
The only one brunch exists in the minimal model, which does not admit the Vainshtein mechanism,
while the new branch appears in the non-minimal model.
It is notable that
the minimal model predicts $\order{(r_{V}/r)^{3}}$, while the non-minimal model does 
$\order{(r_{V}/r)^{3/2}}$.

\section{Summary and Discussion}
\label{sec_summary}

We have studied the asymptotic behavior of the space-time around the relativistic star 
in the dRGT massive gravity with the flat reference metric.
We have explicitly shown that the Vainshtein mechanism does not work in the minimal model,
which is consistent with the previous theoretical analysis~\cite{Renaux-Petel2014a}.
Remarkably, we have found that the modification terms become relevant even inside 
the relativistic star, and thus, that the modification of gravity becomes reasonable not only 
outside the star but also inside the star.
Using the same analysis method,
we have considered the non-minimal model of the dRGT massive gravity.
We have derived the fourth-order algebraic equation based on the several approximations 
and demonstrated the solutions which suggest the non-minimal model has the relativistic 
star solutions with and without the Vainshtein mechanism.

The condition that the Vainshtein mechanism works or not gives a definite difference 
in the equation of motion.
The modification terms to the Einstein equation are integrated into $m_{0}^{2} I_{\mu\nu}$, 
and $\chi(r)$ characterizes $I_{\mu\nu}$.
Because $I_{\mu\nu}$ contains the third order terms of $\chi(r)$ in Eq.~\eqref{Yint2},
the condition $\chi \sim \order{1}$ implies that the modification terms is of $\order{m_{0}^{2}}$,
and on the other hand, the condition $\chi\sim\order{1/m_0}$ predicts that 
the modification includes the term of $\order{1/m_0}$ in general.
The former case shows that the modifications to the equation of motion can be ignored, 
and the latter case shows that the essential contributions from the modifications 
arise in the modified TOV equation.
Therefore, the absence of the Vainshtein mechanism drastically changes the mass-radius 
relation of the relativistic star.
In our previous work~\cite{Katsuragawa2016a}, 
we have obtained the mass-radius relations for the neutron star and quark star 
in the minimal model of the dRGT massive gravity,
which display significant differences from those in the general relativity.
From the above discussion, 
we can understand that the lack of the Vainshtein mechanism in the minimal model have 
produced the differences
because the TOV equations receive the non-negligible modifications.

A couple of comments and discussion on prospects regarding what we have elucidated 
in the present paper are as follows: 
The brunch including the Vainshtein mechanism allows us to impose the ordinary boundary condition,
where we connect the external solution with the Schwarzschild space-time, around the relativistic star.
Therefore, we can compute the mass-radius relation even in the non-minimal model of 
the dRGT massive gravity
based on the techniques which had been established in our previous work.
Although we might face another difficulty to solve the fourth-order equation of $\chi(r)$,
we can solve the modified TOV equation with the arbitrary EoS.
However, the Vainshtein mechanism may result in almost the same mass-radius relation as that in the general relativity. 

Regarding the two branches in the non-minimal model,
we have not constrained the parameter regions to obtain the realistic solution of $\chi(r)$
although we have discussed the leading order and deviation from the Minkowski space-time.
Concerning the relativistic star solution, 
we should evaluate the parameters as well as the mass-radius relation.
Because we have derived the fundamental equations,
we could discuss the particular combination of the parameters to simplify the equation but 
obtain the physical solutions,
which we will address in our future works.


\section*{Acknowledgments}
This work is supported by the Nagoya University Program for Leading Graduate Schools funded 
by the Ministry of Education of the Japanese Government under the program number N01 (M.Y.) 
and by International Postdoctoral Exchange Fellowship Program at Central China Normal University 
and Project funded by China Postdoctoral Science Foundation 2018M632895 (T.K.).
This work is also supported by MINECO (Spain), FIS2016-76363-P, and by project 2017 SGR247 (AGAUR, Catalonia) (S.D.O.)
and by MEXT KAKENHI Grant-in-Aid for Scientific Research on
Innovative Areas ``Cosmic Acceleration'' No. 15H05890 (S.N.) and
the JSPS Grant-in-Aid for Scientific Research (C) No. 18K03615
(S.N.).

\bibliography{library.bib}

\end{document}